\documentclass[aps,pre,showkeys,showpacs,floatfix]{revtex4} 
\usepackage{graphicx}
\usepackage[ansinew]{inputenc}
\usepackage[tbtags]{amsmath}
\usepackage{amssymb}
\usepackage{float}

\newcommand{\beq}{\begin{equation}}
\newcommand{\eeq}{\end{equation}}

\begin{document}
\title{Entropies in case of continuous time}
\author{Detlef Holstein}
\affiliation{Max Planck Institute for the Physics of Complex Systems,
N\"othnitzer Str.\ 38, 01187 Dresden, Germany}
\email{holstein@pks.mpg.de}

\date{\today}
\begin{abstract}
Information theory on a time-discrete setting in the framework of 
time series analysis is generalized to the time-continuous case.
Considerations of the Roessler and Lorenz dynamics as well as the
Ornstein-Uhlenbeck process yield for time-continuous entropies a new 
possibility for the distinction of chaos and noise.
In the deterministic case an upper threshold of the joint uncertainty 
in the limit of infinitely high sampling rate can be found and the 
entropy rate can be calculated as a usual time derivative of the 
entropy. 
In a three-dimensional representation the dependence of the joint entropy 
$H(\epsilon, \tau, t)$ on space resolution, discretization 
time step length and uncertainty-assessed time is shown in a unified 
manner.
Hence the dimension and the Kolmogorov-Sinai entropy rate 
of any dynamics can be read out as limit cases from one single graph.
\end{abstract}

%

\maketitle

\section{Introduction}
Uncertainty of the outcome of random variables $X$ is usually 
evaluated by Shannon entropies \cite{cover91}
\beq
\label{eq:shannonentropy}
H(X)=-\sum_{i=1}^N P(x_i)\ln P(x_i)\; ,
\eeq
where $x_i$ are the possible realizations of $X$. 
Relaxing the restrictions in the underlying Khinchin or Fadeev axioms 
it is possible to introduce the family of Renyi entropies \cite{renyi61}
\beq
\label{eq:renyientrop}
H^{(q)}(X)=\frac{1}{1-q}\ln \sum_{i=1}^N P(x_i)^q \; ,
\eeq
which in the case of Renyi order $q=2$ are quite often used for 
estimation of entropies via the correlation sum \cite{grassproc83}.
In the case of time series analysis it is a common 
task to evaluate the uncertainty of $m$ random variables belonging
to $m$ successive time steps \cite{kantzschreiber04}. 
The corresponding joint entropy in the 
suitable notational representation reads
\beq
H_m^{(q)}(\epsilon, \tau)=\frac{1}{1-q}\ln \left[\sum_{i_1=1}^{M_1(\epsilon)} ...
\sum_{i_m=1}^{M_m(\epsilon)} {(P_{i_1,...,i_m}(\epsilon, \tau))}^q \right] \; .
\eeq
In this expression $\epsilon$ is the space resolution and $\tau$ is the 
step length of the time discretization.
After having decided for a suitable Renyi order, it is often omitted.
Now the finite-m-entropy rate can be defined as 
\beq
\label{eq:finitementropyrate}
h_m(\epsilon, \tau):= \frac{H_{1|m}(\epsilon, \tau)}{\tau}
:=\frac{H_{m+1}(\epsilon, \tau)-H_m(\epsilon, \tau)}{\tau} \; .
\eeq 
This quantity is a special conditional entropy per time step length.
Since obviously this entropy rate is obtained from a quotient 
of differences, a time-continuous formulation of the relationships 
of entropic quantities should naturally also be at hand.
The idea of connection of the entropic quantities by derivatives 
is also supported in \cite{crutch03}, where however, the time 
discretization step length $\tau$ is still kept finite.
The questions of consequences of variable time step length $\tau$ 
and especially the limit of $\tau \rightarrow 0$ for entropic 
quantities are addressed in this paper and it is
intended to give theoretical insights into the structure
of the relationships behind such entropic quantities. 
With those efforts this paper tries to contribute to a symmetrization 
of the treatment of entropies concerning  
space, where continuity is already realized in formulas, and time.

First, the line of thought from joint Shannon entropies to the 
Kolmogorov-Sinai entropy rate is outlined in sec.~\ref{sec:epstau}, 
staying quite close to the paper of Gaspard and Wang \cite{gaspard93}, 
the central paper behind this work. The limit of infinitesimal 
time step length $\tau$ is performed after having performed the 
limit of infinite time $t=m\tau$. 
Theoretically, sec.~\ref{sec:timecontinf} contains the central aim of 
this paper. It is the generalization of sec.~\ref{sec:epstau} with 
omission of the limit of infinite uncertainty-assessed time $t$, concentrating 
on the limit of infinitesimal time discretization step length $\tau$.
In sec.~\ref{sec:examplecalctimecontentrop} the presented ideas are tested
numerically for the examples of the Roessler, Lorenz and Ornstein-Uhlenbeck
dynamics with the detection of qualitative discrepancies of deterministic 
and stochastic dynamics.
Sec.~\ref{sec:conlus} concludes the results of this paper.

\section{From Shannon entropy to Kolmogorov-Sinai entropy rate}
\label{sec:epstau}
The starting point as given in \cite{gaspard93}
is the joint Shannon entropy
\beq
\label{eq:HPtaut}
H({\cal{A}}, \tau, t)=H_{m:=t/\tau}({\cal{A}}, \tau)=-\sum _{\omega_0, ..., \omega_{m-1}}
P(\omega_0, ..., \omega_{m-1} ) \ln P(\omega_0, ..., \omega_{m-1} ) \; .
\eeq
The symbol $\tau$, being the time step length, appears on the 
right side only implicitly as the time between successive realizations
and the partition ${\cal{A}}$ appears on the right side implicitly 
in the range of values $\omega_k$, which corresponds to $i$ in 
eq.~(\ref{eq:shannonentropy}), can take.
In the following it is assumed that the partition for $t_1$ 
is consistent with the partition for $t_2 > t_1$.
It is possible to define the rate
\beq
\tilde{h}({\cal{A}}, \tau, t):=\frac{H({\cal{A}}, \tau, t)}{t} \; .
\eeq
The entropy per unit time with respect to partition ${\cal{A}}$
is then defined 
as the limit
\beq
\tilde{h}({\cal{A}}, \tau)
:=\lim_{t \rightarrow \infty}\tilde{h}({\cal{A}}, \tau, t) \; .
\eeq
Starting from eq.~(\ref{eq:HPtaut}) it is also possible to define another 
rate
\beq
\label{eq:anotherrate}
h({\cal{A}}, \tau, t):=
\frac{H({\cal{A}}, \tau, t+\tau)-H({\cal{A}}, \tau, t)}{\tau}\; ,
\eeq
and the corresponding limit
\beq
\label{eq:partitentroprate}
h({\cal{A}}, \tau):=\lim_{t \rightarrow \infty} h({\cal{A}}, \tau, t) \; .
\eeq
In general it holds
\beq
\tilde{h}({\cal{A}}, \tau, t) \neq h({\cal{A}}, \tau, t)
\quad , \quad \mbox{but} \qquad 
\tilde{h}({\cal{A}}, \tau) = h({\cal{A}}, \tau) \; .
\eeq
The $\epsilon\tau$-entropy rate is derived by
\beq
\label{eq:epsiltauentropratefrompartit}
h(\epsilon, \tau)= \inf_{{\cal{A}}: diam(A_i) \leq \epsilon} 
h({\cal{A}}, \tau) \; .
\eeq
On the other hand, having obtained $H_m(\epsilon, \tau)$ from 
$H_m({\cal{A}}, \tau)$, e.g., again via infimum
\beq
H_m(\epsilon, \tau)= \inf_{{\cal{A}}: diam(A_i) \leq \epsilon} 
H_m({\cal{A}}, \tau) \; ,
\eeq
by the limit of infinite time from the conditional entropy rate
also $\epsilon\tau$-entropy rates can be obtained \cite{cencini00}:
\beq
\label{eq:epsiltauentropratefromepsil}
h_{\infty}(\epsilon, \tau)
 = \lim_{m \rightarrow \infty}h_m(\epsilon, \tau)
 = \lim_{m \rightarrow \infty}
\frac{1}{\tau}[H_{m+1}(\epsilon, \tau)-H_m(\epsilon, \tau)] \; .
\eeq
Depending on the state space resolution $\epsilon$ and time resolution $\tau$,
$h(\epsilon, \tau)$ and $h_{\infty}(\epsilon, \tau)$ 
both are the uncertainty per time step of the immediate future time step 
ahead if infinite time conditioning is imposed.
The Kolmogorov-Sinai entropy rate for processes in continuous time 
is obtained from
\beq
\label{eq:ksentropfromentropyrate}
h_{KS}=\lim_{\epsilon \rightarrow 0, \tau \rightarrow 0}
h_{\infty}(\epsilon, \tau)
=\lim_{\epsilon \rightarrow 0, \tau \rightarrow 0}h(\epsilon, \tau) \; .
\eeq

\section{Time-continuous information theory}
\label{sec:timecontinf}
In the former section the succession of limit procedures for accessing 
the KS entropy rate was redisplayed.
Starting with the same formula (\ref{eq:HPtaut}), it is a naturally arising 
question what happens if the limits of $t\rightarrow \infty$ and 
$m\rightarrow \infty$ are not performed, but instead the limit 
$\tau \rightarrow 0$ is investigated \cite{holstdiss07}. This does not 
lead to dynamical invariants, but instead to prediction-relevant 
quantities of information theory in the time-continuous case, 
because prediction deals with finite-time conditioning.

The partition dependence of eq.~(\ref{eq:HPtaut}) or a similar 
resolution $(\epsilon)$ dependence is suppressed in notation in this
section, since the time aspects should be pointed out, and it remains 
the entropy $H(\tau, t)$. 
Since $H_{m=1}(\tau)$ is the uncertainty of the realization of one 
random variable, it should be independent of $\tau$. This results in
\beq
\label{eq:entropyindepoftau}
H(\tau, t=\tau)=\mbox{const}\quad \mbox{for} \quad \tau > 0 \; ,
\eeq
where a double $\tau$ - dependence on the left side of the equation 
leads surprisingly to $\tau$-independence of the whole expression.

For fixed $t$ now the limit $\tau \rightarrow 0$ has to be carried out.
If the limit exists, then
\beq
\label{eq:Hfctoftaueqzeroandt}
H(\tau = 0, t)\equiv \lim_{\tau \rightarrow 0}H(\tau, t)
\eeq
is the time-continuous joint entropy. Otherwise a time-continuous treatment 
of the joint uncertainty is not possible for the process at hand.
Taking on the other hand the other involved variable $t$ equal to zero
it holds
\beq
\label{eq:Htautequivnull}
H(\tau, t=0) \equiv 0 \; .
\eeq
Hence it is inferred
\beq
H(\tau = 0, t=0) = 0 \; .
\eeq
From $H_m(\tau)=H(\tau, t=m\tau)$ being (not necessarily strong)
monotonously increasing in $m$ it is inferred that also in the limit of
$\tau \rightarrow 0$ the entropy $H(\tau=0, t)$ is 
(not necessarily strong) monotonously increasing in t.
An interesting question is the finiteness of 
$H(\tau=0, t)$ for finite uncertainty-assessed time t
for various process classes, which will be answered numerically by treating 
examples in sec. \ref{sec:examplecalctimecontentrop} shown in fig.'s 
\ref{fig:roesslerentroplinearxtime10tomin2chaosc5k7_t40},
\ref{fig:roesslerentroplinearztime10tomin2chaosc5k7_t40},
\ref{fig:lorenzentropieslinearxtime10tomin2}
and
\ref{fig:ou3D_Hvst}.
For fixed $t>0$ the performed limit causes
\beq
\label{eq:mtoinfty}
\lim_{\tau \rightarrow 0}m=\lim_{\tau \rightarrow 0}\frac{t}{\tau}=\infty \; .
\eeq
From eqs.~(\ref{eq:HPtaut}) and (\ref{eq:mtoinfty}) it has to be inferred 
that $H(\tau = 0, t)$ as the uncertainty of the whole path 
has finally to be understood in terms of 
path integral-type quantities, where the initial and
final states are not fixed. \\

Eq.~(\ref{eq:finitementropyrate}) in notation suitable 
for the purpose of a time-continuous formulation 
or eq.~(\ref{eq:anotherrate}) without partition dependence
read
\beq
\label{eq:htaut}
h(\tau, t) = \frac{H(\tau, t+\tau)-H(\tau, t)}{\tau} \; .
\eeq
If the following limit exists, then
\beq
\label{eq:hvontaugleichnullundt}
h(\tau =0, t) \equiv \lim_{\tau \rightarrow 0} h(\tau, t)
\eeq
is the finite time entropy rate in the time-continuous case.
The discrepancy from the usual differentiation is that the function to be 
differentiated {\em{depends explicitly}} on the parameter of the 
differentiation. This is unusual, but not untreatable.
In case of existence of $H(\tau = 0, t)$ in eq.~(\ref{eq:Hfctoftaueqzeroandt})
for the corresponding derivative it holds that
\beq
\label{eq:usualderivative}
h(\tau = 0, t)=\lim_{\Delta t \rightarrow 0}
\frac{H(0,t+\Delta t)-H(0, t)}{\Delta t} \; ,
\eeq
i.e., $h(0,t)$ from eq.~(\ref{eq:hvontaugleichnullundt})
is obtained as a usual derivative from $H(0,t)$.
$h(\tau=\Delta t, t)$ is found to be a perturbation of the usual quotient of 
differences $\frac{H(0,t+\Delta t)-H(0, t)}{\Delta t}$, but the limit
of $\Delta t \rightarrow 0$ is the same in both cases.
From eqs.~(\ref{eq:htaut}), (\ref{eq:Htautequivnull}) and (\ref{eq:entropyindepoftau}) it is derived
\beq
\tau \cdot h(\tau, t=0)=H(\tau, t=\tau)=\mbox{const} \; .
\eeq
One concludes trivially that for arbitrary dynamics
\beq
\label{eq:divergenceofhzerozero}
\lim_{\tau \rightarrow 0} h(\tau, t=0)
= \infty \quad \mbox{if} \quad H(\tau, t=\tau)>0 \; .
\eeq
For deterministic dynamics the behaviour of the entropy rate $h$ 
will be shown in fig.~\ref{fig:roesslerconddreidimchaos5k7}.
It should be mentioned that the same treatment as here for the first 
derivative can of course also be carried through for higher derivatives
\cite{holstdiss07}.

Eq.~(\ref{eq:htaut}) inverted and iterated leads with
\begin{align}
H(\tau, t=m\tau)
& =(h(\tau, t=(m-1)\tau)+ ...+ h(\tau, 0)) \cdot \tau 
\end{align}
in the time-continuous limit to a representation 
of the joint entropy via integral:
\begin{align}
\label{eq:timecontentasintegral}
H(\tau=0, [0,t]) 
& \quad \;\; \equiv \quad \;\; H(\tau=0, t) \nonumber \\
& \quad \;\; = \quad \;\; \lim_{\tau \rightarrow 0}[\sum_{k=0}^{m-1}\tau \cdot
h(\tau, t=k\tau)] \nonumber \\
& \stackrel{\;\; m\tau \, = \, m'\tau'}{=} 
\lim_{\tau \rightarrow 0} \lim_{\tau' \rightarrow 0} 
[ \sum_{k=0}^{m'-1}\tau' \cdot h(\tau, t=k\tau')]\nonumber \\
& \quad \;\; = \quad \;\; \lim_{\tau \rightarrow 0}\int_0^t du \, 
h(\tau, u) \nonumber \\
& \quad \;\; = \quad \;\; \int_0^t du \, h(\tau=0, u) \; .
\end{align}
Interchangeability of the limit procedures differentiation
and integration is needed in the final step.

\section{Numerical calculations for time-continuous entropies}
\label{sec:examplecalctimecontentrop}
\subsection{Roessler system (deterministic chaotic dynamics)}
\label{sec:zeitkontroessler}
The Roessler system is given by
\begin{align}
\dot{x} &= -y-z \; , \nonumber \\
\dot{y} &= x+ay \; ,\nonumber \\
\dot{z} &= b+z(x-c) \; .
\end{align}
The parameters are chosen as $a=b=0.2$ and $c=5.7$.
The quadratic term '$zx$' is the only nonlinearity.
The largest Lyapunov exponent of the Roessler attractor is 
$\lambda \approx 0.07$
and the fractal dimension is $D^{(2)} = 1.99 \pm 0.07$.

\begin{figure}
\begin{center}
\includegraphics[angle=0, width=12cm]{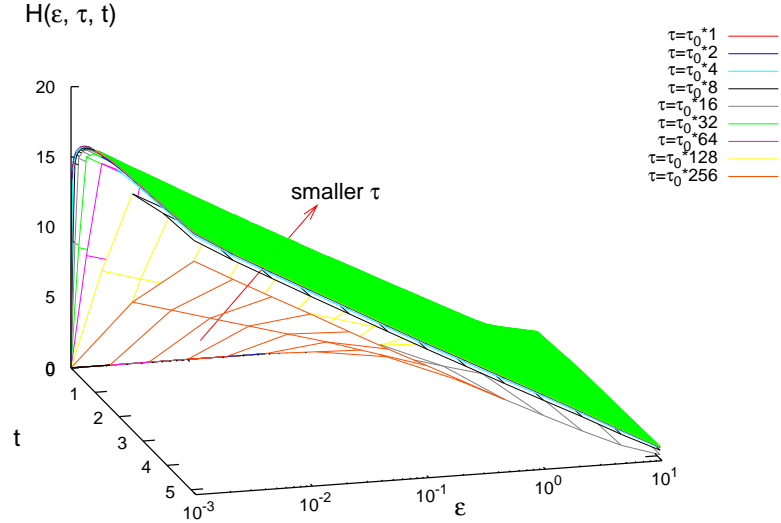}
\end{center}
\vglue -0.8cm
\caption[]{\small\label{fig:roesslerjointdreidimchaosc5k7} 
Joint entropies of the Roessler system  as a function 
of the time $t$ and the resolution $\epsilon$ for various 
$\tau=\tau_0\cdot 2^n\, ; \, n=0,...,8$ ($\tau_0=0.01$)
for the {\em{x-coordinate}} of the Roessler dynamics; $3\cdot 10^6$ 
data points in the generated time series; 
$t=5$ corresponds to approximately 3 circulations in the attractor.} 
\end{figure}
\begin{figure}
\begin{center}
\includegraphics[angle=0, width=12cm]{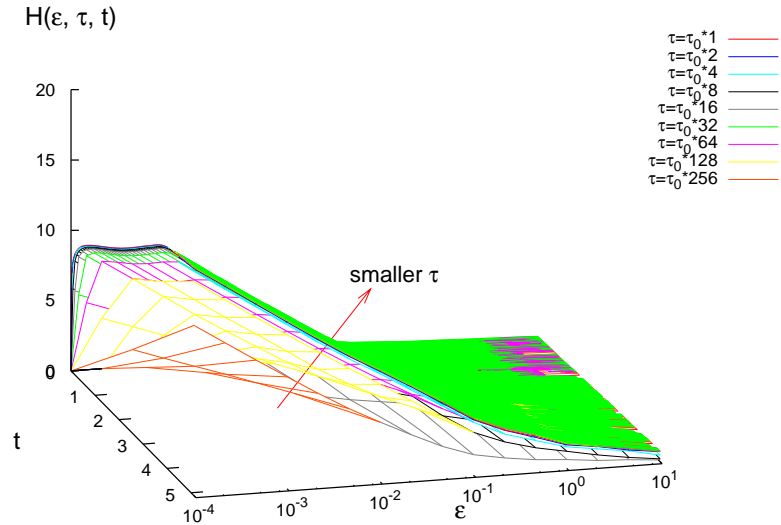}
\end{center}
\vglue -0.6cm
\caption[]{\small\label{fig:roesslerjointzkompdreidimchaosc5k7epsext}
Joint entropies of the Roessler system  as a function of the time $t$
and the resolution $\epsilon$ for various $\tau=\tau_0\cdot 2^n\, ; \, 
n=0,...,8$ ($\tau_0=0.01$)
for the {\em{z-coordinate}} of the Roessler dynamics;
$3\cdot 10^6$ data points.} 
\end{figure}

In fig.~\ref{fig:roesslerjointdreidimchaosc5k7} the joint entropy 
of the x-coordinate of the Roessler system is given as a function of 
resolution $\epsilon$, time step length $\tau$ and 
uncertainty-assessed time $t$.
Convergence of the joint entropy for decreasing time step length 
$\tau$ can be seen.
The joint entropy $H$ depends logarithmically on the resolution $\epsilon$,
from which the dimension is obtained as a slope according to 
\beq
\label{eq:dimensionfromentropy}
D_m^{(q)}=- \lim_{\epsilon \rightarrow 0}
\frac{H_m^{(q)}(\epsilon)}{\ln \epsilon}
\eeq
(\cite{schusterjust05}, p.106) with a value as predicted.

It is possible to see in 
fig.~\ref{fig:roesslerjointzkompdreidimchaosc5k7epsext}
(in comparison with fig.~\ref{fig:roesslerjointdreidimchaosc5k7}
one smaller decade of resolutions is shown)
that the z-coordinate of the Roessler attractor 
carries a different uncertainty for the same values of
($\epsilon, \tau, t$) compared with the x-coordinate.
Nevertheless the dimension of the attractor is captured also by the 
joint entropy of the z-coordinate.

\begin{figure}
\begin{center}
\includegraphics[angle=0, width=10.9cm]{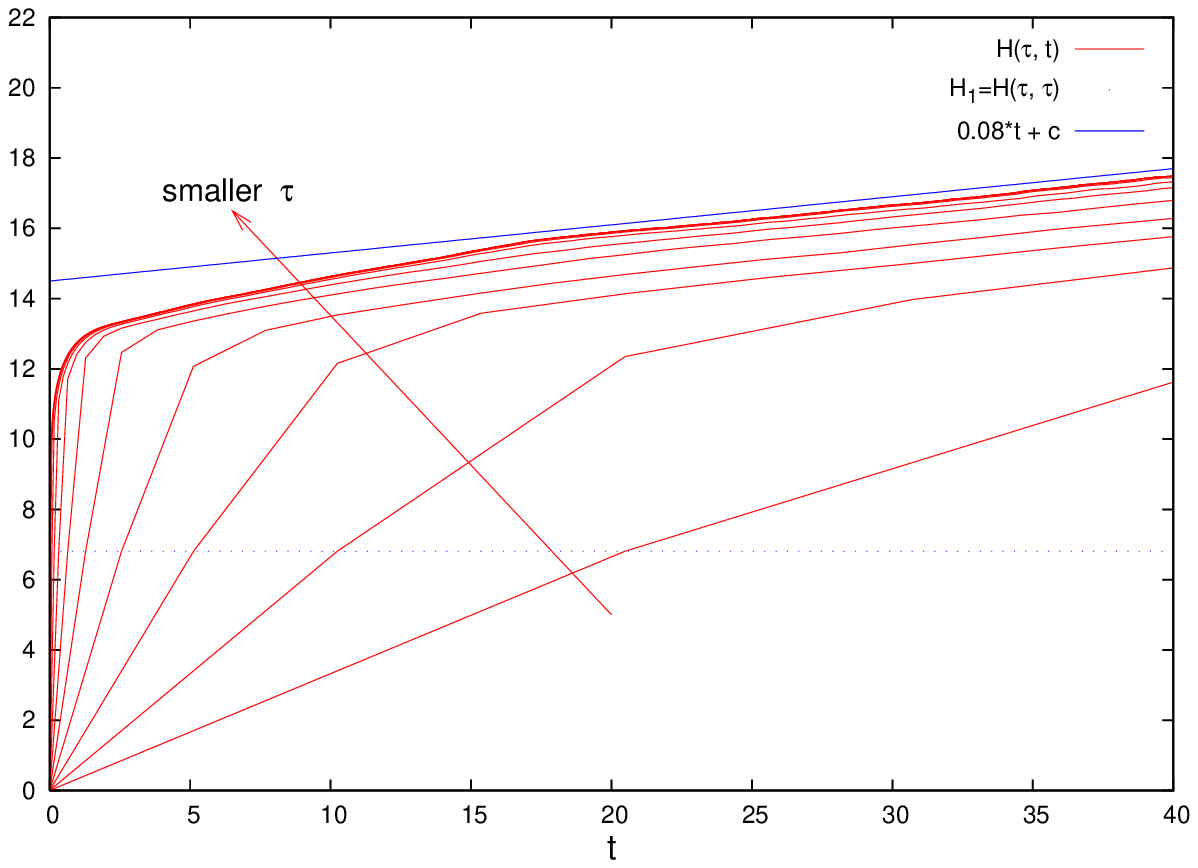}
\end{center}
\vglue -0.2cm
\caption[]{\small\label{fig:roesslerentroplinearxtime10tomin2chaosc5k7_t40}
Joint entropy of the x-coordinate of the Roessler system for fixed 
resolution $\epsilon=10^{-2}$.
With respect to time $t$ it is an extended slice of 
fig.~\ref{fig:roesslerjointdreidimchaosc5k7} for fixed resolution $\epsilon$.
}
\end{figure}

In fig.~\ref{fig:roesslerentroplinearxtime10tomin2chaosc5k7_t40}
the slice of fig.~\ref{fig:roesslerjointdreidimchaosc5k7} for 
resolution $\epsilon = 10^{-2}$ with extended time $t$ is shown.
It is found that asymptotically in $t$ the joint uncertainty of the 
Roessler system increases linearly. The measured slope for 
smallest $\tau$ in between $t=20$ and $t=40$ is 0.08.
It estimates the KS entropy rate
\beq
\label{eq:kolmsinentr}
h_{KS}=\lim_{\epsilon\rightarrow 0}\lim_{\tau\rightarrow 0}
\lim_{t \rightarrow \infty}h(\epsilon, \tau, t) \; .
\eeq
The deviation from the expected value 0.07 can be found in the fact
that still too small $t$ or too large $\epsilon$ are used for the estimation.
It can be concluded from the small slope that the uncertainty of the 
first few points is larger than the uncertainty of a rather long motion 
in the attractor given the first few points.

\begin{figure}
\begin{center}
\includegraphics[angle=0, width=10.9cm]{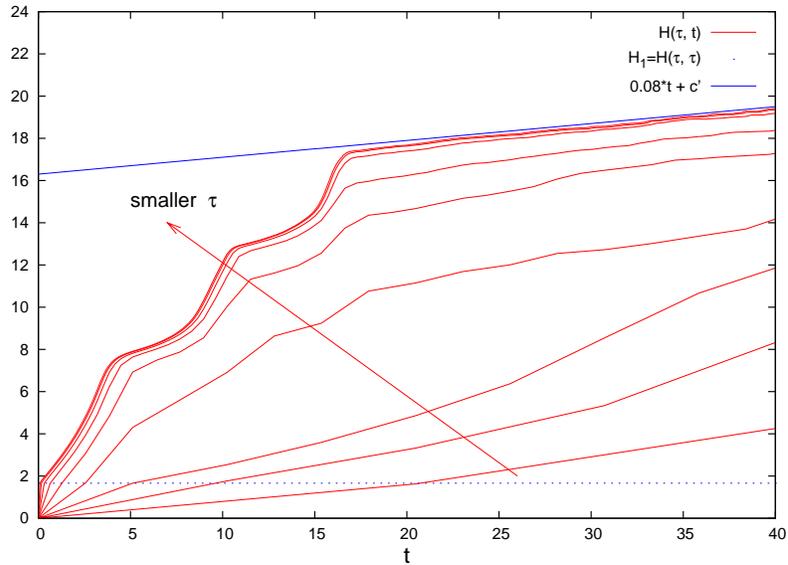}
\end{center}
\vglue -0.2cm
\caption[]{\small \label{fig:roesslerentroplinearztime10tomin2chaosc5k7_t40}
Joint entropy of the z-coordinate of the Roessler system for fixed 
resolution $\epsilon=10^{-2}$. With respect to time $t$ it is an 
extended slice of fig.~\ref{fig:roesslerjointzkompdreidimchaosc5k7epsext}
for fixed resolution $\epsilon$.
} 
\end{figure}

In fig.~\ref{fig:roesslerentroplinearztime10tomin2chaosc5k7_t40}
the slice of fig.~\ref{fig:roesslerjointzkompdreidimchaosc5k7epsext} 
for resolution $\epsilon = 10^{-2}$ with extended time $t$ is shown.
Compared to fig.~\ref{fig:roesslerentroplinearxtime10tomin2chaosc5k7_t40}
a much smaller initial uncertainty $H_1$ is observed for the
z-coordinate. This is understood from the high probability 
of the z-coordinate to stay at zero.
Furthermore in contrast to the x-coordinate a wave-like structure 
of the joint entropy for smaller time $t$ is observed for the z-coordinate. 
A non-monotonous entropy rate $h$ as a function of $t$ has to be inferred.
This unintuitive signature was robustly reproduced under various 
parameter values and confidence in this result is caused by the fact 
that asymptotically for smallest available $\tau$ the slope of 0.08
is found, which serves as a correct estimation of the KS entropy
also from the z-coordinate of the Roessler system. 
Since also (from fig.~\ref{fig:roesslerjointzkompdreidimchaosc5k7epsext})
the dimension  of the Roessler dynamics is correctly extractable 
the result should not be a numerical artefact.
Nevertheless it is not expected that this should be true physics in the 
sense of increasing uncertainty by enlarged conditioning. 
According to Shannon entropies, for which additional conditioning 
cannot increase uncertainty, this behaviour is forbidden.
A possible explanation of this behaviour can be assigned to 
effects of Renyi order $q=2$ (cmp. eq.~(\ref{eq:renyientrop})), 
because in this case it seems unproven that increasing 
conditioning necessarily reduces the resulting entropy.
With this reasoning the example carries the potential of tearing 
apart the interpretation of uncertainty from $q\neq 1$ - entropies.
Another possible explanation that the wave-like structure corresponds
to a finite sample effect being responsible for non-ergodicity has to 
be treated as rather improbable, because also for smallest resolution 
(largest $\epsilon$) it was not possible to detect a qualitative change 
in the sense of smoothing of the wave-like structure of the joint entropy 
under drastical enlargement of the length of the dataset.

With consideration of the fact that the values of $h_{KS}$ and $D$ were 
estimated for finite large $t$ and finite small $\epsilon$ 
instead of in the true limit it can be concluded that 
the same values are obtained from the x- and the z-coordinate 
of the chaotic Roessler system in the limit cases, 
in principle in accordance with the theorem of Takens \cite{takens81}.
On the other hand, it can be clearly seen that for finite time $t$ 
and finite resolution $\epsilon$
the joint uncertainty is allowed to behave qualitatively different
for different observables as the x- and the z-coordinate 
of the chaotic Roessler system.

\begin{figure}
\begin{center}
\includegraphics[angle=0,
width=13cm]{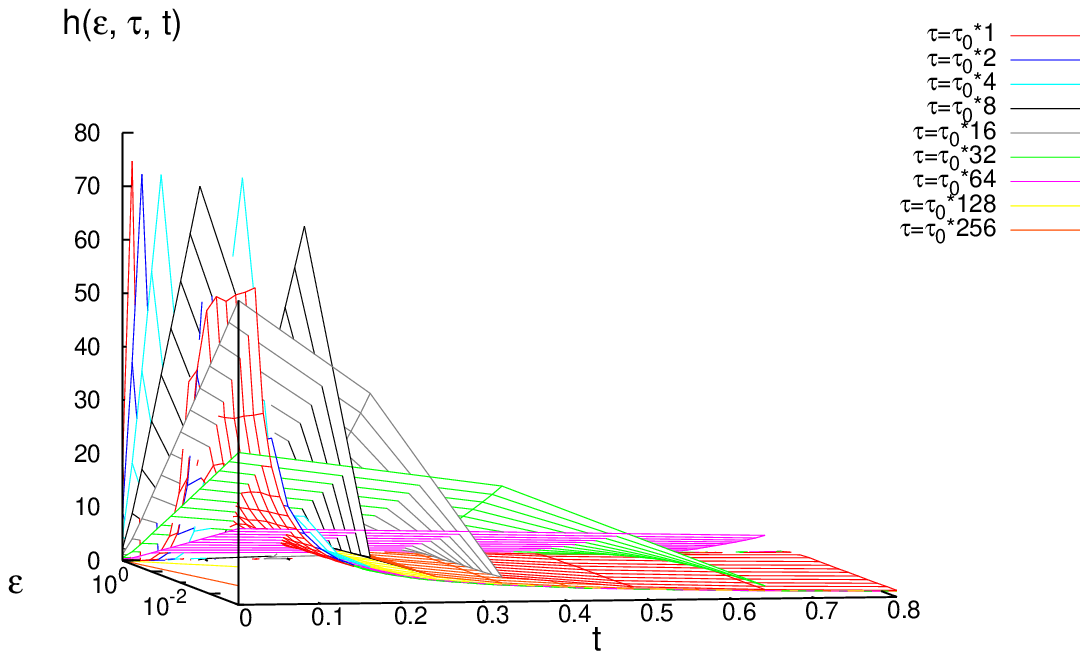}
\vglue -1.5cm
\includegraphics[angle=0,
width=13cm]{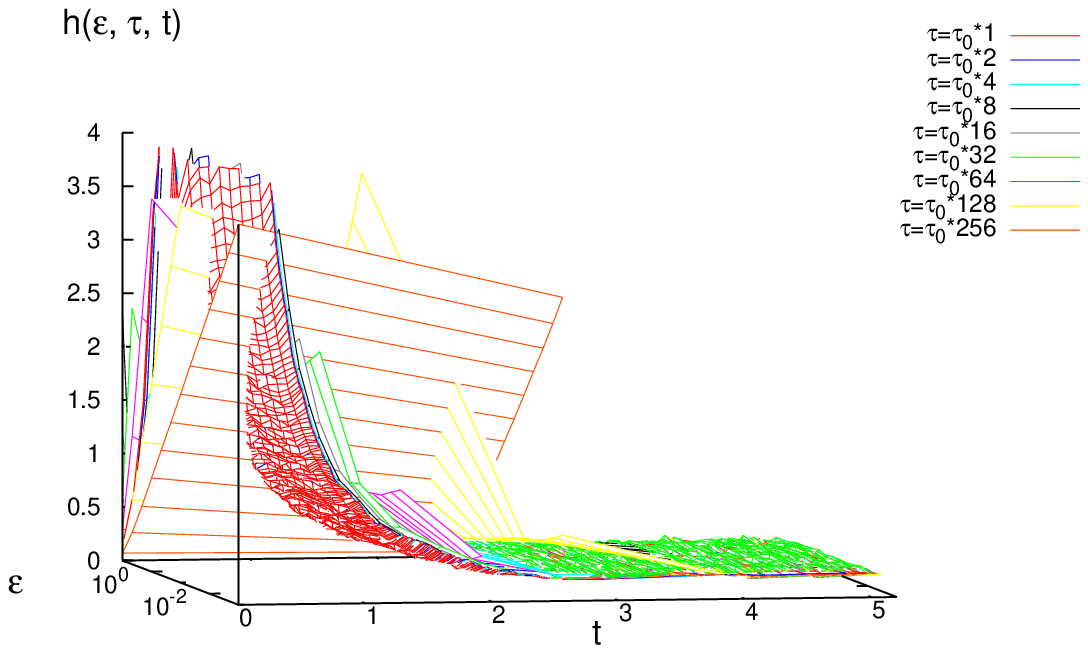}
\end{center}
\vglue -1.5cm
\caption[]{\small\label{fig:roesslerconddreidimchaos5k7} Entropy rate 
as a function of resolution $\epsilon$, uncertainty-assessed time $t$ 
and varying time discretization interval 
$\tau$ for the x-coordinate of the Roessler system with parameters 
for chaotic dynamics.
Upper panel: Focus on resolution dependence of divergent behaviour
for small time $t$.
Lower panel: Focus on larger times $t$ for the 
Kolmogorov-Sinai entropy rate.
} 
\end{figure}

In fig.~\ref{fig:roesslerconddreidimchaos5k7} the entropy rate of the 
Roessler system is shown.
For every $\tau$ the entropy rate as a function of $t$ and $\epsilon$
is represented by a surface. Those surfaces are interleaved 
for different $\tau$. Depending on the ranges shown along the axes 
different results become apperent:
With the example of the Roessler system 
even for deterministic dynamics a true divergence of $h(\tau =0, t=0)$
can be observed for sufficiently small $\epsilon$ seen in particular 
in the upper panel of fig.~\ref{fig:roesslerconddreidimchaos5k7}.
The trivial argument of eq.~(\ref{eq:divergenceofhzerozero})
already gave a hint for such behaviour.
It is possible to see that $h(\epsilon, \tau, t=0)$ for finite $\tau$
increases logarithmically in $\epsilon$ with a slope depending on $\tau$.
In the lower panel for larger $t$ the entropy rate indicates a 
small finite value in the front right corner of the plot, 
which for sufficiently large $t$ approximates the KS entropy rate.

\subsection{Lorenz system (deterministic chaotic dynamics)}
The implemented discretized equations for the Lorenz dynamics are
\begin{align}
x_{n+1} &= x_n+\sigma (-x_n + y_n)\Delta t \; , \nonumber \\
y_{n+1} &= y_n+(-x_nz_n+rx_n-y_n)\Delta t \; , \nonumber \\
z_{n+1} &= z_n+(x_ny_n-bz_n)\Delta t 
\end{align}
with the usual parameters
\beq
r=28.0\; , \qquad \sigma=10.0\; , \qquad b=\frac{8}{3}\; .
\eeq
The fractal dimension of the Lorenz attractor is about 
$D^{(2)} \approx 2.06$ 
and the largest Lyapunov exponent is $\lambda \approx 0.91$.

\begin{figure}[h]
\begin{center}
\includegraphics[angle=0, width=12cm]{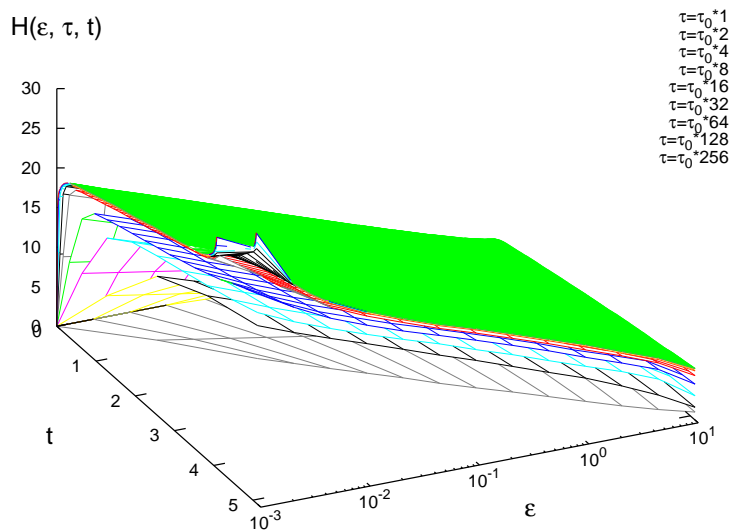}
\end{center}
\caption[]{\small\label{fig:lorenzjointdreidim} Joint entropy of the 
Lorenz system as a function 
of the time $t$ and the resolution $\epsilon$ for various 
$\tau=\tau_0\cdot 2^n\, ; \, n=0,...,8$ ($\tau_0=0.01$)
for the x-coordinate of the Lorenz dynamics; $3\cdot 10^6$ 
data points.} 
\end{figure}
In fig.~\ref{fig:lorenzjointdreidim} the joint entropy of the 
Lorenz system is shown as a function of the resolution $\epsilon$, 
time step length $\tau$ and uncertainty-assessed time $t$. 
As for the Roessler system it is possible to 
see the convergence of the joint entropy $H$ for decreasing time step 
length $\tau$ and a logarithmic dependence of the joint entropy
on the resolution $\epsilon$ with a slope giving the expected 
dimension in the limit of infinitesimal small $\epsilon$.

\begin{figure}
\begin{center}
\includegraphics[angle=0, width=11cm]{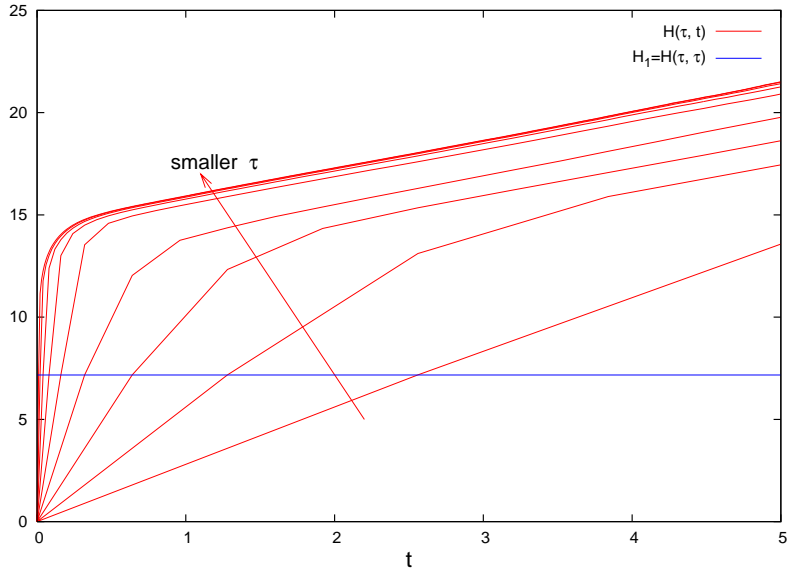}
\end{center}
\caption[]{\small\label{fig:lorenzentropieslinearxtime10tomin2} 
Joint entropy of the Lorenz dynamics for 
resolution $\epsilon = 10^{-2}$ (Slice of fig.~\ref{fig:lorenzjointdreidim});
$3\cdot 10^6$ data points; 
$\tau_0 = 0.01$.
} 
\end{figure}

In fig.~\ref{fig:lorenzentropieslinearxtime10tomin2} a slice of 
fig.~\ref{fig:lorenzjointdreidim} for fixed resolution is shown.
An asymptotically linear behaviour in time
can be found. The fact, that the slope is steeper than that of the
chaotic Roessler attractor 
of fig.~\ref{fig:roesslerentroplinearxtime10tomin2chaosc5k7_t40}
is in accordance with the larger number of nonlinear terms and
the known largest Lyapunov exponents.
The slightly larger slope compared to the known KS entropy of the 
Lorenz system can be explained with estimation at rather large 
$\epsilon$.
The appearance of finite sample fluctuations in entropy estimation 
in fig.~\ref{fig:lorenzjointdreidim}
for small $\epsilon$, small $\tau$ and large $t$ under the same 
estimation conditions sooner in the Lorenz-case than in the 
Roessler-case is in accordance with the enhanced uncertainty $H$ of the
Lorenz dynamics.

Furthermore fig.~\ref{fig:lorenzentropieslinearxtime10tomin2}
indicates that the joint entropies
for different finite $\tau$-values do not coincide for 
asymptotically large times $t$
even though this cannot finally be proven in a plot for finite $t$.
The slopes seem to reach the same value in all cases 
as already observed in 
fig.~\ref{fig:roesslerentroplinearxtime10tomin2chaosc5k7_t40}
for the x-coordinate of the Roessler dynamics, and hence the 
estimation of the KS entropy rate is rather independent 
of the choice of $\tau$, i.e., in the case of deterministic
dynamics the limit with respect to $\tau$ in 
eq.~(\ref{eq:ksentropfromentropyrate}) is not so 
important.
On the other hand, the behaviour of $H(\tau=0, t)$ for finite 
rather small $t$ 
is better resolved for smaller $\tau$, and this in general is rather
important with respect to prediction.

\subsection{Ornstein-Uhlenbeck process (linear stochastic dynamics)}
The Ornstein-Uhlenbeck process
\beq
\dot{X}_t=-\alpha X_t+\sqrt{D}\, \dot{W}_t
\eeq 
can be numerically implemented by its discretisation
\beq
X_{n+1}=(1-\alpha \Delta t)X_n+ \sqrt{D}\sqrt{\Delta t}\eta_{n+1}\; , \quad 
\eta_{n+1}\sim {\mathcal{N}}(0,1)\; \mbox{iid}\; .
\eeq

\begin{figure}
\begin{center}
\includegraphics[angle=0, width=14cm]{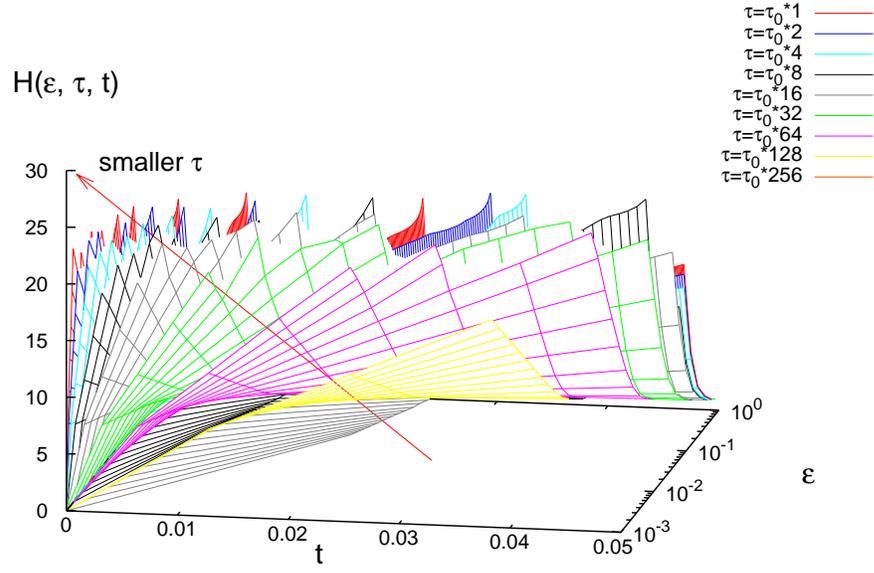}
\end{center}
\caption[]{\small\label{fig:ou3D_Hvst} 
Joint entropy of the Ornstein-Uhlenbeck process. Parameters:
$\alpha=0.025$; $\tau_0=10^{-4}$; 150000 data points.}
\end{figure}

In fig.~\ref{fig:ou3D_Hvst} the numerical analysis 
of the Ornstein-Uhlenbeck process indicates 
the non-existence of the limit 
$\lim_{\tau \rightarrow 0} H(\epsilon, \tau, t)$ 
for sufficiently small $\epsilon$ and finite $t$ except at $t=0$
for time- and amplitude-continuous stochastic dynamics with 
continuous trajectories.
Whereas in the deterministic case, e.g. for the Lorenz dynamics,
the value of the KS entropy rate was approximately seen as a slope 
at sufficiently small finite $\epsilon$ also for finite $\tau$,
in the time- and amplitude-continuous stochastic case the slope 
of $H$ with respect to $t$ 
asymptotically in $t$ for sufficiently small finite $\epsilon$
is seen in fig.~\ref{fig:ou3D_Hvst} to be a function of $\tau$.
Furthermore it was possible to show numerically that for sufficiently small 
$\epsilon$ and non-zero finite fixed time $t$ the entropy rate 
$h(\epsilon, \tau, t>0)$ diverges logarithmically with decreasing 
$\tau \rightarrow 0$ in the example of Ornstein-Uhlenbeck dynamics.
Consistently, fig.~\ref{fig:ou3D_Hvst} supports $h_{KS} = \infty$ 
for this stochastic dynamics, interestingly already from the 
$\tau$-behaviour.
It should be mentioned here that 
for $\tau > 0$ eq.~(\ref{eq:htaut})
can be seen as a naturally given regularization of the unavailable 
eq.~(\ref{eq:usualderivative}) in the stochastic case.

The author is aware of the fact that the result concerning the 
$\tau$-dependence contradicts 
(\cite{gaspard93}, p.322) according to which processes 
with continuous realizations are said to have 
$\tau$-independent $(\epsilon, \tau)$-entropies $h(\epsilon, \tau)$ 
per unit time, which corresponds to 
the suggestion of finite $\lim_{\tau \rightarrow 0} H(\epsilon, \tau, t)$.
The argument for this behaviour was essentially that for finite $\epsilon$
a finite crossing time of underlying boxes of the partition 
leads to finite uncertainty for continuous trajectories 
(\cite{gaspard93}, p.317).
On the other hand, an explanation for the numerically found behaviour
in fig.~\ref{fig:ou3D_Hvst} could be
that infinite uncertainty arises from the trajectories close to 
the border of boxes of partitioning, where in the limit of $\tau \to 0$ 
infinitely often moving back and forth between boxes in finite time $t$
is in principle possible for the Ornstein-Uhlenbeck process. 
This would create an infinite and hence dominant contribution for an 
averaged uncertainty calculation.

\begin{figure}
\begin{center}
\includegraphics[angle=0, width=14cm
]{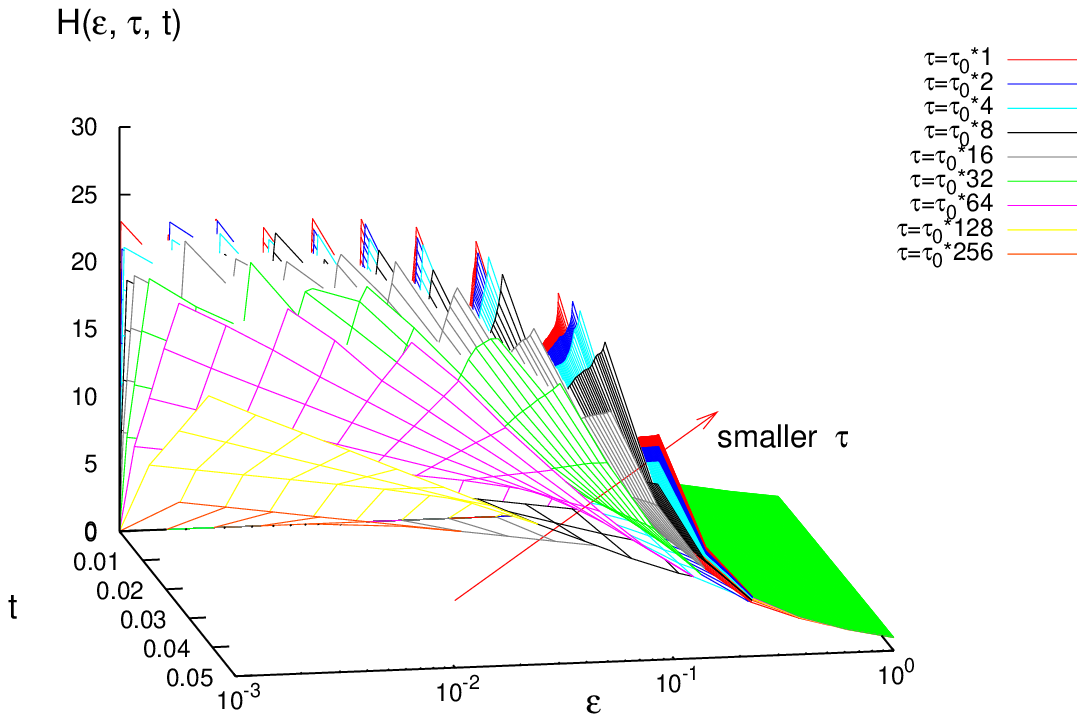}
\end{center}
\caption[]{\small\label{fig:ou3D_Hvseps} The same plot as in 
fig.~\ref{fig:ou3D_Hvst} from another point of view.}
\end{figure}

From fig.~\ref{fig:ou3D_Hvseps} it is obtained that in the transition 
regime of large $\epsilon$ the behaviour of non-existing  
$\lim_{\tau \rightarrow 0} H(\epsilon, \tau, t)$ is suppressed.
The plots do not allow for the decision of the question if
the transition from finite to infinite occurs at the resolution
of the size of the system or at a smaller value for $\epsilon$.
According to eq.~(\ref{eq:dimensionfromentropy})
fig.~\ref{fig:ou3D_Hvseps} yields finite dimensions
for finite $\tau$ increasing in $t$ for stochastic dynamics, 
even though the limit 
$\epsilon \rightarrow 0$ can of course not be seen.
In the limit of $\tau \rightarrow 0$, what is effectively like 
the limit of $m \rightarrow \infty$,
fig.~\ref{fig:ou3D_Hvseps} suggests an infinite dimension
for stochastic dynamics.

The results of this section are qualitatively robust against 
changes of the parameter $\alpha$ of the Ornstein-Uhlenbeck process 
and hence qualitatively robust against changes of the autocorrelation 
of the process.

\section{Conclusion}
\label{sec:conlus}
The central objects of investigation in this paper are the entropy
$H^{(q)}(\epsilon, \tau, t)$ and the entropy rate $h^{(q)}(\epsilon, \tau, t)$
as a function of the resolution $\epsilon$,
the time discretization step length $\tau$ and the uncertainty-assessed time
$t$ with numerical access to the case of Renyi order $q=2$.
Special focus is laid on the analysis of the behaviour of these 
quantities for varying $\tau$ and especially on the time-continuous limit
$\tau \rightarrow 0$ while $t$ is kept finite.
In case of existence, in the time-continuous limit the entropy 
rates can be understood as usual time-derivatives of entropies.
However, for finite $\tau$, in consequence of the explicit 
$\tau$-dependence of entropies, discrepancies from the usual 
quotient of differences are present.

Numerically the analysis of time- and space-continuous dynamics 
was carried out with the Roessler and Lorenz system 
for the deterministic case and with the Ornstein-Uhlenbeck process
for the stochastic case. Qualitative discrepancies of deterministic 
and stochastic dynamics were found.
In the deterministic case the uncertainty $H(\epsilon, \tau=0, t)$ is finite 
for all finite $\epsilon$ and $t$. The entropy rate $h(\epsilon, \tau=0, t)$ 
becomes constant for large $t$ and sufficiently small $\epsilon$.
In the stochastic case the uncertainty $H(\epsilon, \tau=0,t)$ 
is infinite for all $t>0$ if $\epsilon$ is below some threshold. 
Only for $\epsilon$ above the threshold it is finite 
and then effectively like in the deterministic case.
The stochastic result is qualitatively independent of the value of the width 
of the input noise as long as the width is not exactly zero, corresponding to 
determinism and introducing the qualitative change.
This behaviour of entropies in the limit $\tau \rightarrow 0$ 
seems to offer a new possibility for distinction of 
chaos and noise (see also \cite{cencini00}).

The convergence of $H$ in the limit $\tau \rightarrow 0$
in the deterministic case can be interpreted as a saturation 
of the gain from higher sampling rates, 
such that a threshold sampling rate can be postulated,
above which almost nothing more can be learned, i.e., 
in the deterministic case the continuous 
limit can be well approximated by discrete sampling with sufficiently 
high sampling rate. A criterion for an optimal sampling
rate could be formulated.

In the limit case $\epsilon \rightarrow 0$
the partial derivative of the joint entropy $H$ with respect
to $\ln \epsilon$ yields (minus) the dimension and if furthermore 
the limit $t \rightarrow \infty$ is carried out, the partial 
derivative of $H$ with respect to $t$ yields the KS entropy rate. 
All information concerning entropy rates (including the KS entropy rate) 
and dimensions of the dynamics can be extracted from {\em{one single}} plot
of $H^{(q)}(\epsilon, \tau, t)$. 
In the deterministic case known values were verified in examples.
For time- and amplitude-continuous stochastic dynamics it was seen that 
the partial derivative of $H$ with respect to minus $\ln \epsilon$ 
becomes infinite for $t \to \infty$ or $\tau \to 0$
and the partial derivative of H with respect to $t$ becomes infinite 
for $\epsilon \to 0$ or $\tau \to 0$.

According to usual rules the total differential of the 
joint entropy $H$ can be written as
\begin{align}
dH = \frac{\partial H}{\partial t} \, dt
+ \frac{\partial H}{\partial \ln \epsilon} \, d\ln \epsilon \; .
\end{align}
For sufficiently small $\epsilon$ and sufficiently large $t$ this becomes
\beq
\label{eq:differentialofentropy}
dH = h_{KS} \,dt -D \,d\ln \epsilon \; .
\eeq
It is possible to see that directional derivatives 
of $H$ in general mix the properties of dimension and entropy rate.
For integral quantities
$H(\epsilon, t) \approx h_{KS} \, t - D \log \epsilon
+\operatorname{const}$ \;
was suggested in (\cite{grassberger91}, p.529), from which 
eq.~(\ref{eq:differentialofentropy}) can be derived, but 
a prescription for the calculation of the constant is not given.
Since eq.~(\ref{eq:differentialofentropy}) avoids the offset problems
it is a slightly reduced and hence preferable representation of the 
relationship of the involved quantities.

Concerning the limit cases $t \to \infty$ and $\epsilon \to 0$ with the
determination of dynamical invariants, which is discussed in the 
literature, e.g. see (\cite{farmer82}, p.1321), the time-continuous case 
treated in this paper gives a new unified representation of 
singly known results.
On the other hand, at least of the same importance, the case of finite
time $t$ is of strong interest, if questions concerning optimal finite 
time conditioning with respect to prediction are addressed. 
A finite resolution $\epsilon$ is needed for optimality of 
local prediction methods. 
With finite $t$ and $\epsilon$, the limit $\tau \to 0$ is of interest for 
the determination of the informational characteristic of the dynamics.

A non-monotonous entropy rate for Renyi order $q=2$, i.e., a conditional
entropy, which does not monotonously decrease in the length of conditioning,
is numerically found for the z-coordinate of the Roessler system
in fig.~\ref{fig:roesslerentroplinearztime10tomin2chaosc5k7_t40}. 
This could be interpreted as a hint for problems of the interpretation
of uncertainty for entropies with $q\neq 1$ in maximal generality.

\end{document}